\documentclass[journal]{IEEEtran}
\usepackage{amsmath}
\usepackage{amsfonts}
\usepackage{graphicx}
\usepackage{mathrsfs}
\usepackage{algpseudocode}
\usepackage{subfigure}
\usepackage{bm}
\usepackage{cite}
\usepackage{multicol}
\usepackage{multirow}
\usepackage[colorlinks,            
linkcolor=black,            
anchorcolor=black,            
citecolor=black]{hyperref}

\usepackage{url} 
\usepackage{caption}
\usepackage{booktabs}
\usepackage{color}
\usepackage[ruled,linesnumbered]{algorithm2e}

\begin{document}

\title{Learning to Advise and Learning from Advice in Cooperative Multi-Agent Reinforcement Learning}

\author{\IEEEauthorblockN{Yue Jin$^1$, Shuangqing Wei$^2$, Jian Yuan$^1$, Xudong Zhang$^1$}\\
\IEEEauthorblockA{
$^1$ Department of Electronic Engineering, Tsinghua University, Beijing, China \\
$^2$ School of Electrical Engineering and Computer Science, Louisiana State University, Baton Rouge, USA \\
jiny23@126.com, swei@lsu.edu,
jyuan@tsinghua.edu.cn, zhangxd@tsinghua.edu.cn}
}

% \author{Yue~Jin, \IEEEmembership{Student~Member,~IEEE}, Shuangqing~Wei, \IEEEmembership{Senior Member,~IEEE}, Jian~Yuan, \IEEEmembership{Member,~IEEE},\\
% and Xudong Zhang, \IEEEmembership{Member,~IEEE}
% \thanks{
% \quad This work was supported in part by the National Natural Science Foundation of China under Grant U20B2060.

% Y. Jin, J. Yuan, and X. Zhang are with the Department of Electronic Engineering, Tsinghua University, Beijing, 100084, China
% (e-mail: jiny23@126.com; jyuan@tsinghua.edu.cn; zhangxd@tsinghua.edu.cn ).

% S. Wei is with the School of Electrical Engineering and Computer Science, Louisiana State University, Baton Rouge, LA 70803 USA (e-mail: swei@lsu.edu).}
% }

\maketitle
\IEEEpeerreviewmaketitle

\begin{abstract}
Learning to coordinate is a daunting problem in multi-agent reinforcement learning (MARL). Previous works have explored it from many facets, including cognition between agents, credit assignment, communication, expert demonstration, etc. However, less attention were paid to agents’ decision structure and the hierarchy of coordination. In this paper, we explore the spatiotemporal structure of agents’ decisions and consider the hierarchy of coordination from the perspective of multilevel emergence dynamics, based on which a novel approach, Learning to Advise and Learning from Advice (LALA), is proposed to improve MARL.
% to propel the emergence of coordination in MARL.
Specifically, by distinguishing the hierarchy of coordination, we propose to enhance decision coordination at meso level with an advisor and leverage a policy discriminator to advise agents’ learning at micro level. The advisor learns to aggregate decision information in both spatial and temporal domains and generates coordinated decisions by employing a spatiotemporal dual graph convolutional neural network with a task-oriented objective function. Each agent learns from the advice via a policy generative adversarial learning method where a discriminator distinguishes between the policies of the agent and the advisor and boosts both of them based on its judgement. Experimental results indicate the advantage of LALA over baseline approaches in terms of both learning efficiency and coordination capability. Coordination mechanism is investigated from the perspective of multilevel emergence dynamics and mutual information point of view, which provides a novel perspective and method to analyze and improve MARL algorithms.
\end{abstract}


\section{Introduction}
The emergence of collective intelligence is a fascinating topic. People are intrigued by creating collective artificial intelligence. Deep multi-agent reinforcement learning (MARL) is one of the inspiring achievements in cooperative agents \cite{ liu2021motor, jin2021hierarchical, li2019cooperative, peng2017multiagent, vinyals2019grandmaster } in recent years. 
However, learning to coordinate is still a daunting problem in MARL. 

Many facets have been explored to propel the emergence of coordination and accelerate learning. Learning cognition between agents is a natural idea when it comes to coordination. Fruitful investigation have been conducted along this direction \cite{wen2019probabilistic, jin2020stabilizing, jin2021information}. Credit assignment \cite{sunehag2018value, Rashid2018QMIX, son2019qtran, jakob2018counterfactual} is another critical exploration, whose goal is to learn individual credit in the settings with a sharing reward function. 
Another branch of studies proposed to learn from experts. A typical class of methods is imitation learning, where agents learn from expert demonstrations \cite{song2018multi, yu2019multi, jeon2020scalable, le2017coordinated, zhan2018generative}. 
However, the coordination mechanism is either not investigated or just explored from the perspective of cognition between two agents in these methods.
Communication provides a means of coordinating in a larger range and thereby facilitates coordination. Plenty of studies have been conducted to learn communication content \cite{sukhbaatar2016learning, foerster2016learning}, choose communication agents \cite{jiang2018learning, das2019tarmac} and decide when to communicate \cite{singh2018learning}. Essentially, they involve information sharing mechanism, but seldom consider decision structure. 
Recently, some works \cite{li2021deep,yuan2022GCS} have paid attention to decision execution structure, where agents make decisions based on a learned coordination graph reflecting their decision dependency. 
% However, the sequential execution manner may cause less efficiency than synchronous execution. 
However, the intrinsic spatiotemporal structure of agents’ decisions have not been explored. In addition, these works mentioned above barely involve mutual interactions among agents’ decisions.
% The structure formed by agents is a critical feature emerges along with collective intelligence.

\begin{figure}
 \centering
%  %Requires \usepackage{graphicx}
\includegraphics[width=0.6\columnwidth]{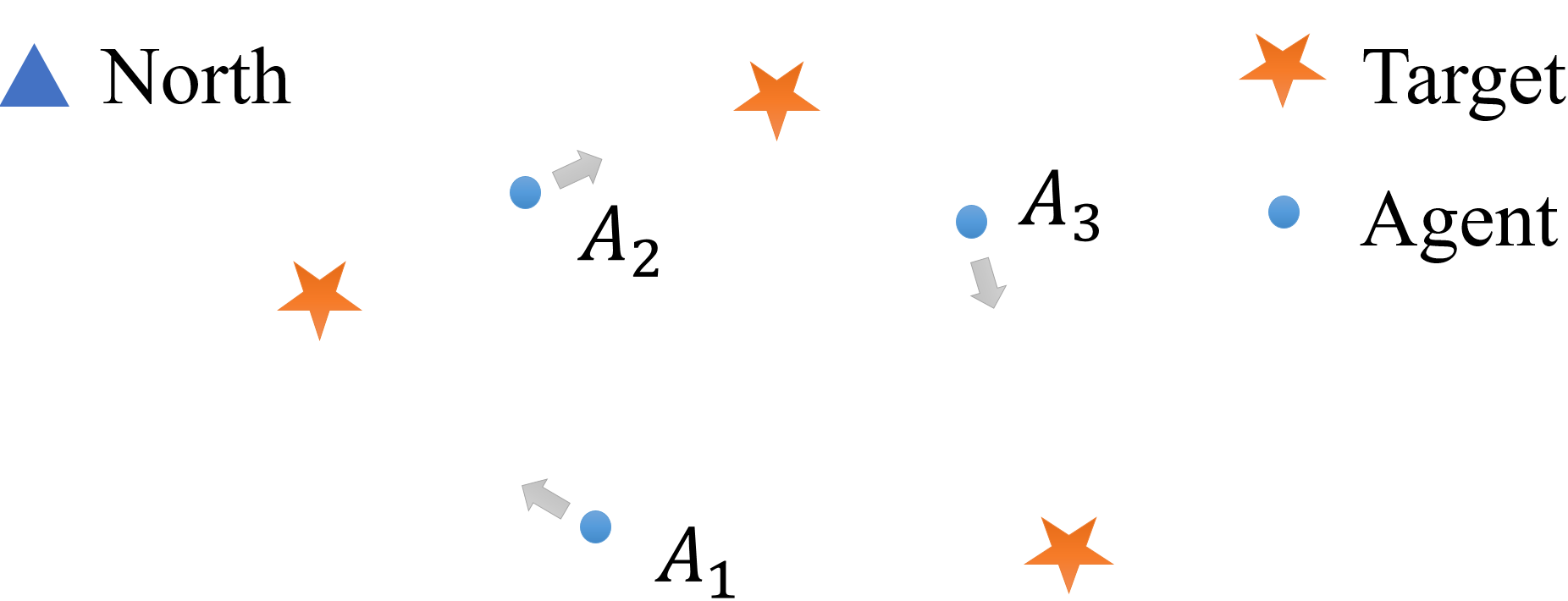}
% \captionsetup{font=footnotesize}
\caption{An illustration of cooperative navigation. }\vspace{-1mm}
\label{env}
\end{figure}

In this paper, we explore the spatiotemporal structure of agents'  decisions and the hierarchy of coordination, and propose a novel algorithm of Learning to Advise and Learning from Advice (LALA) to guide agents' learning in cooperative MARL. Specifically, we consider the hierarchy of coordination from the perspective of multilevel emergence dynamics (MED)\cite{ribeiro2020multilevel, kozlowski2016capturing, klein2000micro, rogov2018urban}, where meso-level interaction plays a vital role in the emergence process of coordination.
Therefore, we propose to enhance decision interplay at meso level via a centralized advisor who leverages the spatiotemporal decision structure of agents to coordinate decisions. Micro-level agents are guided by the advisor on policy learning in addition to their own exploration.
We focus on cooperative multi-agent tasks where agents need to avoid decision conflicts, such as cooperative navigation tasks \cite{lowe2017multi, jin2019efficient, jin2020stabilizing, jin2021information} (shown as Fig.~\ref{env}) where agents need to select different targets and minimize the overall navigation time, and multi-arm robot control tasks \cite{liu2021collaborative, jacak2009conflict, jacak2007heuristic} involving position conflicts. 
Coordination includes resolving decision conflicts and smoothing decisions in these tasks.

Specifically, we employ a spatiotemporal graph to characterize the relations between agents’ decisions in both spatial and temporal domains. We propose a spatiotemporal dual graph neural network (DualGCN) to aggregate agents’ decision information and generate coordinated decision advice. The DualGCN takes the graph structure and agents’ decisions as inputs, and mitigates decision conflicts and discontinuity by distinguishing and clustering decision features of different agents and different time steps, respectively.

A policy-level generative adversarial network (PLGAN) is proposed for agents to learn from the advice. It includes a discriminator that distinguishes between the policies of the agent and the advisor. 
The discriminator’s judgement is utilized to guide agent's policy learning and boost the coordination capability of DualGCN. 
Agent's decision network adopts an information bottleneck \cite{tishby2015deep} based encoder for low-dimensional representation and compression of observations. The learned latent variables are used to infer other agents' actions, and taken as inputs of agent's decision network.
Note that, LALA is developed to help learn to coordinate. Once learned, only the decision network is deployed onto each agent to complete tasks.

We investigate the efficacy and advantage of LALA from multilevel system perspective and mutual information point of view. Specifically, mutual information gain between agent's local observation and other agents' decisions is utilized to examine the cognition between agents at micro level. At meso level, we verify the coordination improvement caused by the decision interaction of DualGCN. Bridging different levels, we investigate if the advice generated from meso-level decision interaction affects the cognition between micro-level agents, and examine how the micro-level cognition and meso-level interaction affect macro-level coordination performance.

The main contributions of this paper are as follows.
\begin{itemize}  %[leftmargin=*]
    \item A LALA approach is proposed to improve MARL and propel coordination, which coordinates agents' decisions based on their spatiotemporal structure and guides agents' learning with the coordinated decision advice. 
    \item  We investigate coordination mechanism from the perspective of multilevel emergence dynamics and mutual information point of view, which provides a novel perspective and method to analyze and improve MARL algorithms.
    \item Extensive experimental results demonstrate the advantage of LALA over baseline MARL algorithms in terms of both learning efficiency and cooperation performance.    
\end{itemize}

%%%%%%%%%%%%%%%%%%%%%%%%%%%%%%%%%%%%%%%%%%%%%%%%%%%%%%%%%%%%%%%%%%%%%%%%
\section{Problem Formulation}
% \subsection{Cooperative Multiagent RL}
We consider a Stochastic Game (SG) with $N$ agents, which involves state $s$, joint action $\bm{a} = (a_1,\cdots, a_N)$, transition probability function $p(s'|s, \bm{a})$, and reward function $r_i(s, \bm{a}), i \in \{1, \cdots, N\}$.
At each time step, each agent executes an action according to its policy $\pi_i$ which maps the state to an action probability. Let $\bm{\pi} = (\pi_1, \cdots, \pi_N)$ denote the joint policy.
% The policy could be either deterministic or stochastic. Accordingly, a deterministic policy denoted as $\pi_i(s): S \rightarrow A$, maps states to actions. 
% A stochastic policy, denoted as  $\pi_i(\cdot|s): S \times A \rightarrow [0,1]$, gives an action probability in each state.

In this work, we focus on solving a kind of fully cooperative tasks where 1) the policy of each agent is stochastic and with discrete action space; 2) a decision conflict emerges when two agents choose the same action.
Formally, $a_i \in \{c_z\}_{z=1}^Z$, where $Z$ is the size of the action set, $c_z$ is the $z$-th element of the action set. The policy of each agent satisfies $\forall z, \pi_i(a_i=c_z|s) \in [0,1]$, and $\sum_{z=1}^Z \pi_i(a_i =c_z|s) = 1$.
When $\arg \max _{c_z} \pi_i(a_i=c_z|s) = \arg \max _{c_z} \pi_j(a_j=c_z|s)$, a conflict occurs, and all the agents cannot get maximal reward.

The goal is to find the optimal policy for each agent so that $\forall \pi_i,  R_i(s^t, \pi_1^*, \cdots, \pi_i^*, \cdots, \pi_N^*)\geq R_i(s^t, \pi_1^*, \cdots, \pi_i, \cdots, \pi_N^*)$, where $\pi_i^*$ denotes the optimal policy of agent $i$, and $R_i (s^t, \bm{\pi}) = E[\sum\nolimits_{\tau=0}^T {{\gamma ^\tau}} {r_i(s^{t+\tau},\bm{a}^{t+\tau} )}]$ denotes the expectation of its cumulative discounted reward.

%%%%%%%%%%%%%%%%%%%%%%%%%%%%%%%%%%%%%%%%%%%%%%%%%%%%%%%%%%%%%%%%%%%%%%%%
\section{PRELIMINARIES}
In this section, we present some preliminary knowledge, including background about MARL and an algorithm used as a baseline in our experiments; background about GCN and a general model; multi-agent imitation learning with GAN and a typical algorithm involved in our baseline methods.

\subsection{MARL}  \label{3.1}
The optimal policy of an SG with unknown transition probability and reward function can be solved by MARL.
An individual Q-function \cite{leibo2017multi, jin2021hierarchical} is defined as follows:
\begin{equation}\label{Q function}
Q_i(s,a_i) = {\mathbb{E}}[\sum\limits_{\tau}  {{\gamma ^{\tau}}} {r_i^{t+\tau} }|s^t=s,a_i^t=a_i ].
\end{equation}
It measures the expected total discounted rewards of agent $i$ who takes action $a_i$ in state $s$ and following $\pi_i$ thereafter.
An extended action-value function is proposed to take other agents' actions into consideration \cite{jin2021hierarchical}, which is defined as
$Q_i^{\pi_{i}}(s^t,\zeta(s_{-i}^t, s_{-i}^{t+1}),a_i^t)$. 
$\zeta$ is an estimation function of other agents' actions, $s_{-i}$ is the part of state regarding other agents.
An approximate Bellman equation for the optimal extended Q-function is derived as:
\begin{equation}\label{Bellman_Q}
\begin{aligned}
&\mathbb{E}_{s_{-i}^{t+1}} Q^*_i(s^t, \zeta(s_{-i}^t, s_{-i}^{t+1}),a^t_i) \approx\\ & \mathbb{E}_{s^{t+1}}
[r_i^{t+1}+\gamma \mathop{\max}\limits_{a^{t+1}_i}Q^*_i(s^{t+1}, \zeta(s_{-i}^t,s_{-i}^{t+1}),a^{t+1}_i) ],\\
\end{aligned}
\end{equation}

\begin{figure}[!t]
    \centering
    \includegraphics[width=0.9\columnwidth]{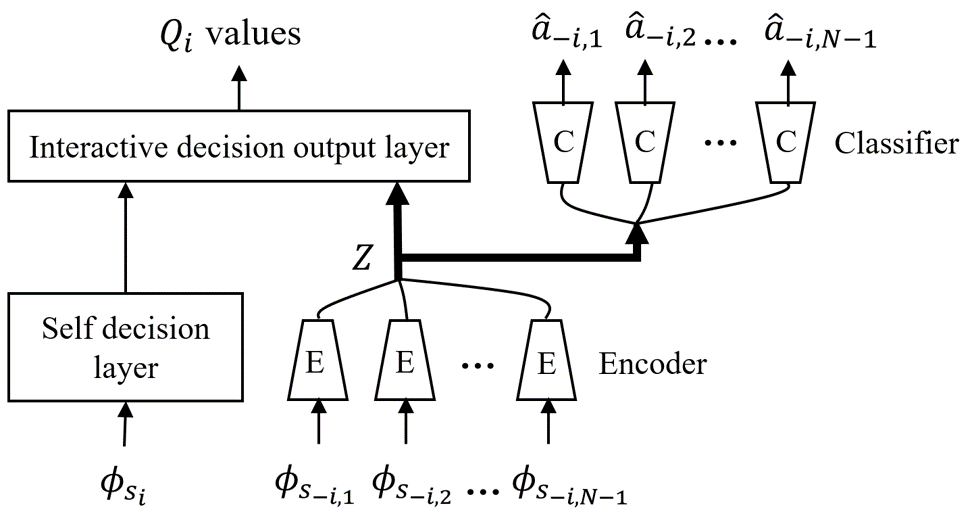}
    % \captionsetup{font=footnotesize}
    \caption{Network architecture diagram of IBORM \cite{jin2021information} .}
    \vspace{-1mm}
    \label{iborm}
\end{figure}

Based on \cite{jin2021hierarchical}, IBORM algorithm \cite{jin2021information} leverages information bottleneck principle to compress the input information of action estimation, while retaining the information regarding other agents' actions. 
The network architecture of IBORM is shown in Fig.~\ref{iborm}.
The loss function for the extended Q-function is given as:
\begin{equation}\label{loss_Q}
\begin{aligned}
    &\mathcal{L}_i^{\theta_i} = \mathbb{E} \left[\left(y_i^t - Q_i^{\theta_i}(s^t, \eta(\phi_{-i, 1}^{t+1}), \cdots, \eta(\phi_{-i, N-1}^{t+1}),a^t_i)\right)^2\right] \\
    & + \sum_{j=1}^{N-1} \left[ \rho_1 \cdot \mathcal{L}_{CE}(\widehat{a}_{-i,j}^t; a_{-i,j}^t) +\rho_2 \cdot I(\phi_{-i,j}^{t+1}; \eta(\phi_{-i, j}^{t+1})) \right.\\
    & \left.- \rho_3 \cdot I(\eta(\phi_{-i, j}^{t+1}); a_{-i,j}^{t})\right] ,
\end{aligned}
\end{equation}
where $\theta_i$ is parameters of the extended Q-function; 
$\phi_{-i, j}^{t+1}=(s_{-i, j}^t, s_{-i, j}^{t+1})$ denotes the states of the $j$-th agent except agent $i$ at two adjacent time steps;
$\eta$ is an encoder to extract the representation relevant to the action of an agent;
$\mathcal{L}_{CE}$ denotes the cross entropy between the estimated action $\widehat{a}_{-i,j}^t=\kappa(\eta(\phi_{-i, j}^{t}))$ and the true action $a_{-i,j}^t$; $\kappa$ is a function that maps the action representation to the estimated action; $I$ denotes mutual information; and $y_i^t = r_i^{t+1}+\gamma \max\nolimits_{a_i^{t+1}} Q_i^{\theta_i}(s^{t+1},\eta(\phi_{-i, 1}^{t+1}), \cdots, \eta(\phi_{-i, N-1}^{t+1}), a^{t+1}_i)$.
Based on the two mutual information terms, IBORM learns optimal representations of other agents' actions and thus helps learn coordinated policies more efficiently.

\subsection{GCN}  \label{3.2}
Graph is an effective tool to represent the relationships between entities in a system, which has been used to represent physical systems \cite{sanchez2018graph}, social networks \cite{wu2020graph},  road networks \cite{li2018diffusion}, protein interaction networks \cite{fout2017protein}, etc.
We denote a graph as $\mathcal{G}=(\mathcal{V}, \mathcal{E})$, where $\mathcal{V}$ is a set of vertices, and $\mathcal{E}$ is a set of edges.
Let $e_{uv} = (u, v)$ denote an edge connecting vertex $u$ and vertex $v$.
The set of neighbors of vertex $v$ is denoted as $\mathcal{N}(v) = \{u|(u, v) \in \mathcal{E}\}$. 

Based on the graph structure and leveraging convolution operators, a GCN can aggregate information within neighborhoods \cite{zhang2019graph, wu2020comprehensive}. 
As multiple graph convolutional layers are stacked, neighbors within multiple skips can be involved in information aggregation. 
A GCN takes the original features of each vertex and the graph structure as input, and can output features for each vertex, each edge, or the overall graph according to the need of a task, such as node classification \cite{hamilton2017inductive}, link prediction \cite{ying2018graph}, or graph classification \cite{zhang2018end}, respectively. 

GraphSAGE \cite{hamilton2017inductive} was proposed as a general aggregation-based graph convolution model.
In the $k$-th convoluational layer, features are aggregated as:
\begin{equation} \label{graphsage}
\begin{aligned}
    & {\bm{h}}_{\mathcal{N}_v}^{k} = \text{AGG}_k (\{ {\bm{h}}_u^{k-1}, \forall u \in \mathcal{N}(v)\})  \\
    & {\bm{h}}_{v}^k = \varphi (g ^k ([{\bm{h}}_v^{k-1}, {\bm{h}}_{\mathcal{N}_v}^k]),
\end{aligned}
\end{equation}
where $\text{AGG}_k$ denotes an aggregation function, e.g., mean or max-pooling function; ${\bm{h}}_v^{k}$ denotes the features of vertex $v$ output by the $k$-th hidden layer; $[\cdot, \cdot]$ denotes concatenated vectors; $g^k$ denotes a linear function, and $\varphi$ denotes a nonlinear activation function.
Specifically, in each convolutional layer, GraphSAGE first aggregates features of the neighbors for each vertex using an aggregation function. Then, it concatenates the aggregated neighbors' features  with the vertex's own features, and uses the concatenated features to compute the output features of this vertex. 

\subsection{GAN in Multi-Agent Imitation Learning} \label{3.3}
A branch of studies in imitation learning leverages GAN to match the data distribution of the learned policy with the distribution of the expert demonstration data.
% It has been proved to be a powerful tool in terms of decreasing compounding error \cite{rajaraman2020toward} in imitation learning.
Some works \cite{song2018multi, yu2019multi, jeon2020scalable} have employed this idea in multi-agent imitation learning.
Typically, MAGAIL \cite{song2018multi} solves polices by optimizing the following objective:
\begin{equation}
\begin{aligned}
    \min\limits_{\bm{\pi}} \max\limits_{D_1, \cdots, D_N} 
    & \mathbb{E}_{\bm{\pi}} [\sum_{i=1}^N \log D_i (s, a_i)] \\
    & + \mathbb{E}_{\bm{\pi}^E} [\sum_{i=1}^N \log (1-D_i (s, a_i))],
\end{aligned}
\end{equation}
where $D_i$ denotes discriminator, $\bm{\pi^E} = (\pi_1^E, \cdots, \pi_N^E)$ denotes the experts' policies.
Each discriminator maps state-action pairs to judging probabilities and thereby distinguishes between state-action pairs collected by agent's policy and the expert demonstrations. 
The policy of each agent plays the role of generator who fools the discriminator by matching its state-action distribution with that of the expert demonstration data.

%%%%%%%%%%%%%%%%%%%%%%%%%%%%%%%%%%%%%%%%%%%%%%%%%%%%%%%%%%%%%%%%%%%%%%%%
\section{Method}
In this section, we propose our proposed LALA method through the lens of MED.
MED is a concept regarding a dynamic process of a system from micro-level individual behaviors and meso-level interplay, to macro-level emergence \cite{ribeiro2020multilevel, kozlowski2016capturing, klein2000micro, rogov2018urban}.
From the perspective of MED, we consider a cooperative multi-agent system as follows: (1) the micro-level behaviors lie in each agent's decision making; (2) the meso-level interplay corresponds to information interaction and decision coordination among agents; (3) the macro-level emergence is reflected by the success of completing a cooperation task.
As an intermediate process, the meso-level interplay plays an important role in the emergence of coordination.
Therefore, we propose LALA to enhance the meso-level decision interplay beyond micro agent scope with an centralized advisor module, and guide agents' learning at micro level. 

\begin{figure*}[!t]
 \centering
%  %Requires \usepackage{graphicx}
\includegraphics[width=2\columnwidth]{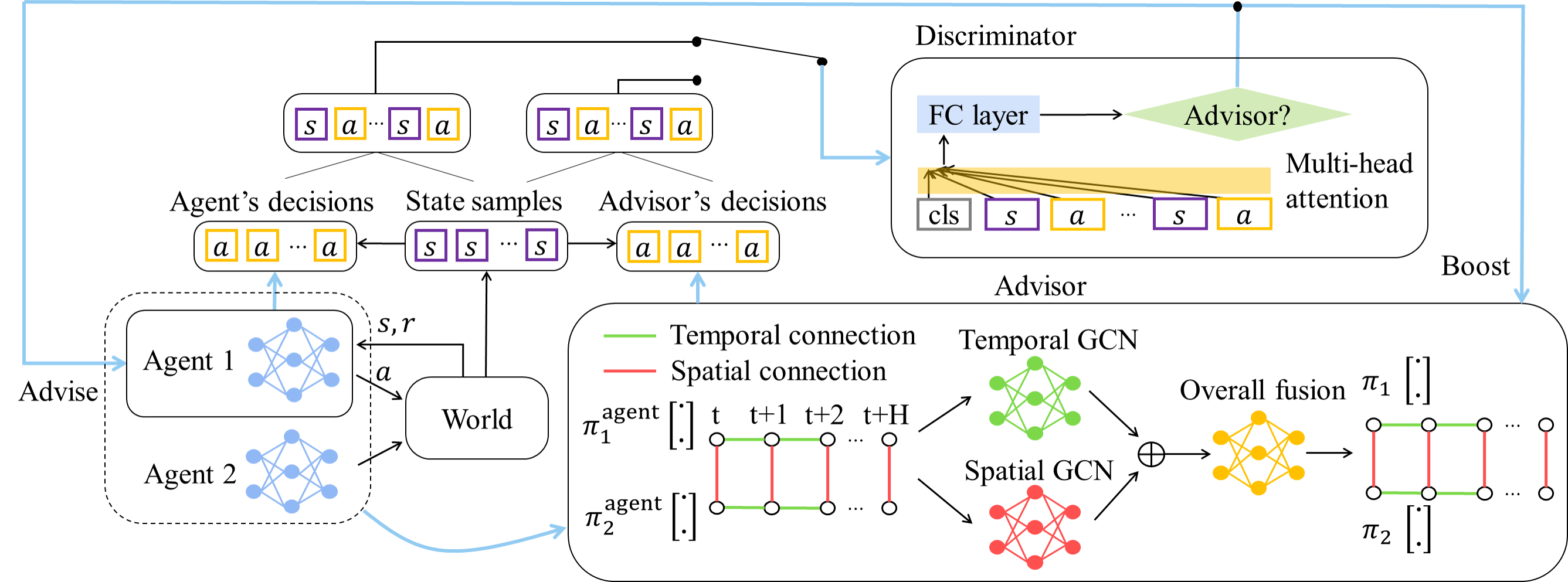}
% \captionsetup{font=footnotesize}
\caption{Illustration of LALA framework. The bottom left part represents MARL agents who interact with the environment. The bottom right part illustrates DualGCN advisor whose input is a  spatiotemporal graph including decisions of all agents. The upper part is a discriminator distinguishing between the state-action sets of the advisor and the agent. The discriminator provides guidance to boost the policies of both the MARL agents and the advisor.}\vspace{-1mm}
\label{LALA_framework}
\end{figure*}

Fig.~\ref{LALA_framework} shows an overall illustration for LALA.
At micro level, each agent makes decisions based on its local observations, and receives reward from the world.
At meso level, a centralized advisor learns to coordinate agents' decisions based on agents' decisions and the spatiotemporal structure. 
At macro level, a discriminator learns to distinguish between the policies of an agent and the advisor.
Based on discriminator's judgement, each agent learns to match with the advised policy, and the advisor learns to boost its advising capability.
The framework is identical to each agent. 
Fig.~\ref{LALA_framework} just takes one agent as an example. Blue arrows indicate the inter-level information flow.

Next, we first present the  advisor (Section~\ref{4.1}), followed by the MARL agents and discriminators (Section~\ref{4.2}), and then an advisor boosting method (Section~\ref{4.3}) based on the discriminators.

% \begin{figure}[!t]
%  \centering
% %  %Requires \usepackage{graphicx}
% \includegraphics[width=0.8\columnwidth]{fig/graph_eg.png}
% \caption{Illustration of the spatiotemporal graph. }\vspace{-1mm}
% \label{graph_eg}
% \end{figure}

\subsection{Meso-Level: Learning to Advise with DualGCN} \label{4.1}
At meso level, an example including two agents ia shown in the bottom right part of Fig.~\ref{LALA_framework}.
We leverage a spatiotemporal graph to characterize the relations between agents’ decisions. 
Thus, the input feature of each vertex is the 
decision made by an agent at a time step.
% , denoted as $\pi_v^{\text{agent}}$.
% DualGCN fuses agents' decision information in both temporal and spatial domains.
% Vertices of the same agent are connected (green edges) along the time dimension. Vertices of different agents at the same time step are connected with each other (red edges).
In this work, we focus on the fully cooperative tasks where agents need to avoid decision conflicts with each other, such as avoiding target selection conflicts in cooperative navigation tasks. 
To reflect the potential conflict between each two agents, vertices of different agents at the same time step are connected with each other in spatial dimension (red edges).
Additionally, considering time continuity required in many control problems, i.e. the decisions at each two adjacent time steps need to be similar, we connect the vertices of the same agents along time dimension (green edges). 
Through the lens of the spatiotemporal graph, the advisor can view the decisions of each agent over the whole system and multiple time slots including the future,
% In contrast, each agent can only make decisions based on its local and current (and historical) observation.
% Therefore, the advisor can
and thus generate better decisions to advise agents.

To generate advice, we propose a spatiotemporal dual GCN (DualGCN), where spatial conflicts and temporal discontinuity are mitigated via decision interaction at meso level.
Specifically, to cope with 
constraints imposed on agents' decisions for maintaining temporal continuity and resolving spatial conflicts, the neighbors of each vertex are divided into two sets, $\mathcal{N}_s(v)$ and $\mathcal{N}_t(v)$. 
DualGCN fuses decision information within $K$-hop spatial and temporal neighborhoods in different manners (weights) and then fuses the overall spatiotemporal information to generate decision advice. 
An abstract illustration is shown in the bottom right part of Fig.~\ref{LALA_framework}.

Formally, in the $k$-th hidden layer, features are aggregated as follows:
\begin{equation} \label{gcn}
\begin{aligned}
    & \bm{m}_s^k = \text{AGG}_k^s (\{ {\bm{h}}_w^{k-1}, \forall w \in \mathcal{N}_s(v)\})  \\
    & \bm{m}_t^k = \text{AGG}_k^t (\{ {\bm{h}}_u^{k-1}, \forall u \in \mathcal{N}_t(v)\})  \\
    & {\bm{h}}_{\mathcal{N}_v}^{k} = f^k ([\bm{m}_s^k, \bm{m}_t^k]) \\
    & {\bm{h}}_{v}^k = \varphi (g ^k ([{\bm{h}}_v^{k-1}, {\bm{h}}_{\mathcal{N}_v}^k]),
\end{aligned}
\end{equation}
where $\text{AGG}_k^s$ and $\text{AGG}_k^t$ denote spatial and temporal aggregation function, respectively; ${\bm{h}}_v^{k}$ is the features of vertex $v$ output by the $k$-th hidden layer; $[\cdot, \cdot]$ denotes concatenating; $f^k$ and $g^k$ are linear functions to be learned; $\varphi$ is a nonlinear activation function.
The number of hidden layers determines how many hops of neighbors each vertex can interact with.
The more hidden layers the DualGCN has, the larger the scope of meso-level decision interactions is.
The aggregating method is similar to GraphSAGE \cite{hamilton2017inductive}.
However, our method considers two types of neighbors, which is not involved in GraphSAGE.

The objective function for DualGCN is designed to mitigate spatial conflicts and temporal discontinuity by distinguishing and clustering the decisions belonging to different agents and time steps, respectively.
Specifically, the goal is to lead the outputs of the vertices adjacent in temporal domain to be similar, and in spatial domain to be distinct.
Therefore, we design a cost function consisting of temporal discontinuity cost and spatial conflict cost by measuring the decision similarity, shown as the first two terms in (\ref{loss_Advisor}).  

Additionally, as a conflict reconciler, the advisor needs to decide which agents should concede and which ones should not.
To generate advice that can improve agents in a gradual manner, the advisor investigates the decision confidence of each agent, and corresponds its decision with the most confident one. 
To this end, we design a max-confidence consistency cost which measures the decision distance between the of the advisor and the most confident agent with Jensen-Shannon divergence.
An indicator function expressing whether the decision is the most confident is given as:
\begin{equation} \label{GCN_label}
    I_{label}(v) = \left\{ \begin{gathered}
                            \begin{aligned}
                            &1 {\text{,   if  }} \forall w \in \mathcal{N}_s(v)\\
                            & \quad \max_a\pi_{v}^{\text{agent}}(a) >\pi_{w}^{\text{agent}} (\arg \max_a\pi_{v}^{\text{agent}}(a)) \\
                            &0 {\text{,   otherwise}} \hfill \\
                            \end{aligned}
                            \end{gathered}  \right. , \\    
\end{equation}
where $\arg \max_a\pi_{v}^{\text{agent}}(a)$ is the preferred action of the agent represented by vertex $v$, and $\max_a\pi_{v}^{\text{agent}}(a)$ is the corresponding decision probability.
As decision probability reflects decision confidence, (\ref{GCN_label}) means that if the preferred action of an agent has the highest possibility among all agents' possibilities on this action, the decision is the most confident. 

The overall cost function of DualGCN advisor is given as:
\begin{equation} \label{loss_Advisor}
\begin{aligned}
&    \mathcal{L}_{Advisor} = \sum_{v \in \mathcal{V}} \left[ \sum _{u \in \mathcal{N}_t(v)} - \log (\sigma(\pi_v^T \pi_u))  \right.\\
&  \left. - \sum _{w \in \mathcal{N}_s(v)} \log(\sigma(-\pi_v^T \pi_w)) + I_{label}(v)\cdot JSD(\pi_v || \pi_v^{\text{agent}})\right] .
\end{aligned}
\end{equation}
where $\sigma$ is the sigmoid function; the first two terms correspond to temporal discontinuity cost and spatial conflict cost; the last term corresponds to the max-confidence consistency cost.

The complete algorithm of learning to advise with DualGCN is shown in Algorithm \ref{alg:GCN}.

\begin{algorithm}[tb] 
  \caption{Learning to Advise with DualGCN}
  \label{alg:GCN}
% \begin{algorithmic} [1]
% \small
   \KwIn {Graph $\mathcal{G}(\mathcal{V}, \mathcal{E})$; vertex features $\{{\pi}_v^{\text{agent}}, \forall v \in \mathcal{V}\}$; state set $\mathcal{B}_i^s$, vertex set $\mathcal{B}_i^v$, 
  depth $K$, training epoch number $E$}
   {\bf Initialize}: Spatial weight matrices $\Omega_k^s$,
  temporal weight matrices $\Omega_k^t$, 
  overall fusion weight matrices $\Omega_k^o$, $ k \in \{1, \cdots,K\}$ \;
%    {\bfseries Output:} Action advice $\pi_v \text{ for each } v \in \mathcal{V}$
   ${\bm{h}}_v^0 \leftarrow {\pi}_v^{\text{agent}}, \forall v \in \mathcal{V}$ \;
  \For {epoch $ = 1 \cdots E $}{
  \For {$k= 1 \cdots K$}{
    \For {$v \in \mathcal{V}$}{
         Compute: ${\bm{h}}_{\mathcal{N}_v}^k = \Omega_k^s \cdot \text{AGG}_k^s (\{ {\bm{h}}_w^{k-1}, \forall w \in \mathcal{N}_s(v)\})+ \Omega_k^t \cdot \text{AGG}_k^t (\{ {\bm{h}}_u^{k-1}, \forall u \in \mathcal{N}_t(v)\})$ and
         ${\bm{h}}_{v}^k = \varphi (\Omega_k^o \cdot \text{CONCAT}({\bm{h}}_v^{k-1}, {\bm{h}}_{\mathcal{N}_v}^k))$ \;
    }
%    ${\bm{h}}_{v}^k \leftarrow ({\bm{h}}_{v}^k+{\bm{h}}_{v\text{min}}^k)/||{\bm{h}}_{v}^k+{\bm{h}}_{v\text{min}}^k||_2, \forall v \in \mathcal{V}$ 
  }
   ${\pi}_v \leftarrow \text{Softmax}({\bm{h}}_{v}^K), \forall v \in \mathcal{V} $ \;
   Update $\Omega_k^s, \Omega_k^t, \Omega_k^o, k \in \{1, \cdots,K\}$ to minimize (\ref{loss_Advisor}) or (\ref{loss_Advisor_Boost}) \;
}
% \end{algorithmic}
\end{algorithm}

%%%%%%%%%%%%%%%%%%%%%%%%%%%%%%%%%%%%%%%%%%%%%%%%%%%%%%%%%%%%%%%%%%%%%%%%
\subsection{Micro-Level: Learning from Advice with PLGAN} \label{4.2}

% To learn from the advice, we propose PLGAN.
% GAN has been developed in imitation learning including multi-agent imitation learning, and achieved promising results.
% However, imitation learning methods require expert demonstration data.
To bridge the information flow between the meso-level DualGCN advisor and micro-level agent's decision network, we propose a macro-level discriminator to distinguish between the advisor and agent policies and help boost them, which is illustrated as the upper part of Fig.~\ref{LALA_framework}. 

To learn from the policy behind the advisor's advice given in the situations encountered by agents, we leverage a policy representation approach analogous to a recent work \cite{jin2021supervised} to learn features of a policy from a set of states and the actions taken by the policy.
The discriminator distinguishes between the agent and advisor based on the policy features.

Specifically, the discriminator consists of a Transformer encoder and a classifier.
The Transformer encoder takes a set of state-action pairs that is structured as a sequence as raw input.
A class token (CLS) \cite{devlin2019bert} is concatenated with the raw input.
The learned embedding of the CLS token is deemed the representation of the policy and then mapped by the classifier to a judging probability. 
Let $\mathcal{B}_i^s$, $\mathcal{A}_i^\pi$, and $\mathcal{A}_i^G$ denote a set of states, actions mapped by the agent's policy, and the advised actions given by the DualGCN advisor, respectively. 
$\mathcal{D}^{\psi_i}$ denotes the discriminator parameterized with $\psi_i$.
The loss function for the discriminator is defined as:
\begin{equation} \label{loss_Disc}
\begin{aligned}
    \mathcal{L}_{Disc, i} = - \mathbb{E}_{\mathcal{B}_i^s} &\left[\log D^{\psi_i}\left(\mathcal{A}_i^G(\bm{\Omega}), \mathcal{B}_i^s\right)\right. \\
    & \left. + \log\left(1-D^{\psi_i}\left(\mathcal{A}_i^\pi(\theta_i), \mathcal{B}_i^s\right)\right) \right],
\end{aligned}
\end{equation}
where $\bm{\Omega}$ denotes parameters of DualGCN, and $\theta_i$ denotes parameters of the policy of agent $i$.
Note that, the advisor is learned in a centralized manner and provides decision advice for all of the agents.
Therefore, $\bm{\Omega}$ is shared by all agents.

At micro level, each agent leverages the corresponding discriminator's judgement to match with the advisor's policy. 
Specifically, each agent learns to fool the discriminator, i.e. maximize the cost function (\ref{loss_Disc}), in addition to learning by its own exploration.
The loss function for each agent is defined as:
\begin{equation} \label{loss_Agent}
    \mathcal{L}_{Agent, i} = \mathcal{L}_i^{\theta_i}+\lambda  \mathbb{E}_{\mathcal{B}_i^s} \log(1-D^{\psi_i}(\mathcal{A}_i^\pi(\theta_i), \mathcal{B}_i^s)), 
\end{equation}
where $\mathcal{L}_{\pi}^{\theta_i}$ is the loss function for policy learning of an MARL algorithm and $\lambda$ is a positive weight that balances agent's active learning and learning from advice. 
Note that, our method is different from imitation learning where reward signal is not available. 
We focus on the case where agents can receive environmental rewards and leverage the advice to improve learning. 

The optimal loss corresponding to (\ref{loss_Disc}) is the Jensen-Shannon divergence between the state-action sets (policy features) of the agent and the advisor.
By minimizing (\ref{loss_Disc}) and (\ref{loss_Agent}), each agent finds a policy under the constraint of matching its policy with the advised policy.
Because the advised policy is coordinated and smoothed, agents can be guided to coordinate.

%%%%%%%%%%%%%%%%%%%%%%%%%%%%%%%%%%%%%%%%%%%%%%%%%%%%%%%%%%%%%%%%%%%%%%%%
\subsection{Meso-Level Boosting via Macro-Level Discriminator} \label{4.3}
In Section~\ref{4.1}, we leverage DualGCN to reconcile decision conflicts and smooth decisions from a global spatiotemporal view, and thereby generate better decisions to advise the MARL. 
Intuitively, when we have two policies and know which is better, we can further boost the better one by exploiting strengths and avoiding weaknesses.
Based on this idea, in this part, we propose to boost the advisor.

% In order to exploit strengths and avoid weaknesses of a policy, we need to evaluate a policy first.
Actually, 
trained with (\ref{loss_Disc}), the discriminator presented in Section~\ref{4.2} can give a higher probability score to the advisor than an agent. This probability score can thus be deemed a metric measuring the extent of advantage that the advisor's policy has over that of an agent. 
Therefore, the discriminator plays the role of an advantage function, which can be 
employed to boost the performance of the advisor's policy.
We introduce an additional term into the original loss (\ref{loss_Advisor}) of DualGCN, which reflects the advantage of the advisor's policy. Thus, the loss function for the advisor is given as:
\begin{equation} \label{loss_Advisor_Boost}
\begin{aligned}
&    \mathcal{L}_{Advisor, boost} = \sum_{v \in \mathcal{V}} \left[ \sum _{u \in \mathcal{N}_t(v)} - \log (\sigma(\pi_v^T \pi_u)) \right.\\
& \left. - \sum _{w \in \mathcal{N}_s(v)} \log(\sigma(-\pi_v^T \pi_w))
 + I_{label}(v)\cdot JSD(\pi_v || \pi_v^{\text{agent}})\right]  \\
& - \mu \sum_{i=1}^N \mathbb{E}_{\mathcal{B}_i^v, \mathcal{B}_i^s} \left[\log D^{\psi_i}(\{\pi_v\}_{v\in \mathcal{B}_i^v}, \mathcal{B}_i^s)\right],
\end{aligned}
\end{equation}
where $\mu$ is a positive weight; $\mathcal{B}_i^s$ denotes a set of states obeying the same sampling distribution as that used to train the discriminator; $\mathcal{B}_i^v$ denotes the vertex index set corresponding to $\mathcal{B}_i^s$.
% We use LALA-Boost to name the LALA method involving advisor boosting.

The complete algorithm of LALA is shown in Algorithm \ref{algo:LALA}.

\begin{algorithm}[tb]
\caption{LALA in MARL}
\label{algo:LALA}
% \begin{algorithmic}[1]
% \raggedright
% \small
    {\bf Initialize}: Parameters $\bm{\Omega}$ for GCN $G^{\bm{\Omega}}$; parameters $\theta_i$ for policy $\pi^{\theta_i}$ and $\psi_i$ for discriminator $D^{\psi_i}$ of each agent; state buffer $\mathcal{D}_i^s$ of each agent; overall decision buffer $\mathcal{D}_\pi$; graph buffer $\mathcal{G}$ \;
\For{episode $=1$ to $C$}{
    %  $\mathcal{D}_\pi \leftarrow$ empty decision buffer
     Receive $s_i^1$ for each agent \;
    \For {$t=1$ to $T$}{
        \For {$i=1$ to $N$}{
             Execute action $a_i^t \sim \pi^{\theta_i}(\cdot|s_i^t)$ \;
             Record state $\mathcal{D}^s_i \leftarrow \mathcal{D}^s_i \cup s_i^t$ \;
             Record decision $\mathcal{D}_\pi \leftarrow \mathcal{D}_\pi \cup \pi^{\theta_i}(\cdot|s_i^t)$ \;
            \If {$|\mathcal{D}^s_i| > \xi$}{
                 Sample state subset $\mathcal{B}_i^s = \{s_i^{\tau_m}\}_{m=1}^M$ from $\mathcal{D}^s_i$ and record vertex index $\mathcal{B}_i^v = \{v_i^{\tau_m}\}_{m=1}^M$ \;
                 Compute decisions $\mathcal{A}_i^{\pi} = \{\pi^{\theta_i} (\cdot|s_i), \forall s_i \in \mathcal{B}_i^s\}$ \;
                 Compute advice $\mathcal{A}_i^G = \{G^{\bm{\Omega}}(v_i), \forall v_i \in \mathcal{B}_i^v\}$ \;
                 Update $\psi_i$ to minimize (\ref{loss_Disc}) \;          
            }
        }
         Receive $r_i^{t+1}, s_i^{t+1}$ for each agent $i$ \;
        \For {$i=1$ to $N$}{
             Update $\theta_i$ to minimize (\ref{loss_Agent}) with $\mathcal{B}_i^s$ \;
        }
    }
     Construct a spatiotemporal graph $\emph{g}$ with all agents' decisions made in the last episode \;
     Record graph $\mathcal{G} \leftarrow \mathcal{G} \cup \emph{g}$ \;
    \For {$i=1$ to $N$}{
         Sample state subset $\mathcal{B}_i^s = \{s_i^{\tau_m}\}_{m=1}^M$ from $\mathcal{D}^s_i$ and record vertex index $\mathcal{B}_i^v = \{v_i^{\tau_m}\}_{m=1}^M$ \;
    }
     Update $\bm{\Omega}$ using Algorithm \ref{alg:GCN} with $\mathcal{G}$, $\mathcal{D}_\pi$, $\{ \mathcal{B}_i^s \}_{i=1}^N$, $\{ \mathcal{B}_i^v \}_{i=1}^N$ \;
}
% \end{algorithmic}
\end{algorithm}
%%%%%%%%%%%%%%%%%%%%%%%%%%%%%%%%%%%%%%%%%%%%%%%%%%%%%%%%%%%%%%%%%%%%%%%%
\section{Experiments}
In this section, we evaluate LALA in multi-agent cooperative navigation task with the same settings used in \cite{ jin2020stabilizing, jin2021information}.
In this task, $N$ agents need to cooperate to reach the same number of targets using the minimum time.
An example containing three agents and targets is illustrated in Fig.~\ref{env}.

The task has a continuous state space and a discrete action space.
Each agent can observe the relative positions of targets and other agents.
At each time step, each agent selects a target and moves a fixed distance ($1 \ \text{m}$) towards it.
Let $a_i \in \{\epsilon\}_{\epsilon=1}^N$ denote the target index selected by an agent. 
Agents are regarded as mass points. 
Potential decision conflicts exist in their target selections.
Environment size is $15\times15~m^2$.
Maximum episode length is $30$ time steps.
Agents are homogeneous. They share a common policy and reward function, which is aligned with \cite{jin2020stabilizing} and \cite{jin2021information}. Reward consists of time penalty $r_{i,step}= -1/L$, conflict penalty $r_{i,conf}=-45/L$, and cooperation award $r_{i,coop}=0.8/L$, where $L$ is the side length of the environment.

We evaluate the performance of each method in the cases with different numbers of targets and agents ($N=7,8,9,$ and $10$).
In training, agents are trained by 20,000 episodes. 
At the beginning of each episode, positions of targets and agents are generated randomly.
After training, we test the trained policies with 1000 episodes.
All experiments are run with three different random seeds.

\subsection{Baselines}
We compare the performance of LALA with the following baselines which include a regular MARL algorithm without an advisor, and three algorithms using advisor but with different advising manners.
The algorithms using advisor are all implemented based on the regular MARL algorithm and they all use the DualGCN advisor.
% Their differences are the manner of leveraging the advice.
% Specifically, the regular MARL algorithm is used in their advisee learning. The advisor learning module is aligned with Section~\ref{4.1}. 
Detailed descriptions of each baseline algorithm are as follows.
% \begin{itemize}
% \item \textbf{Regular learning without advice (No-Advice)} - 
\subsubsection{Regular learning without advice (No-Advice)}
We adopt the algorithm proposed in \cite{jin2021information} as the baseline method without advising. Each agent takes the positions of targets, the current and last positions of other agents, and its last action as the input of its Q-network. 
The loss function for the  Q-network is given as (\ref{loss_Q}).
The Q-network outputs $N$ values that denote the Q-values of $N$ actions. In execution, each agent selects its target according to the maximum Q-values.

% \item \textbf{Knowledge distillation based advising (KDA)} - 
\subsubsection{Knowledge distillation based advising (KDA)}
We compare LALA with the  knowledge distillation \cite{hinton2015distilling, chen2019data, liu2018knowledge} based advising approach where the Q-network with a softmax output layer is regarded as the Student network and the DualGCN used in the advisor learning is regarded as the Teacher network. Instead of leveraging a discriminator to minimize the distribution distance between the policy features of the agent and advisor, KDA directly minimizes the distance between the output of the Student network and the advice given by the Teacher network.
We adopt the Jensen-Shannon divergence to measure the distance, and introduce the distance term into the the original loss function of the No-Advice method. The loss for KDA is defined as:
\begin{equation}
    \mathcal{L}_{KDA} = \mathcal{L}_{\theta}+\lambda JSD(\pi^{\text{agent}}||\pi^{\text{advisor}}),
\end{equation}
where $\mathcal{L}_{\theta}$ is the loss function of the No-Advice method.
$\pi^{\text{agent}}$ is defined as:
\begin{equation} \label{pi_agent}
    \pi^{\text{agent}}=\frac{{\text{exp}} (Q_\theta(s_t,\cdot))}{Z_\theta(s_t)},
\end{equation}
where $Z_\theta(s_t)$ is a normalization factor. 
$\pi^{\text{advisor}}$ is the corresponding advice given by DualGCN.

% \item \textbf{LALA without advisor boost (LALA-NB)} - 
\subsubsection{LALA without advisor boost (LALA-NB)}
To verify the advantage of using advisor boosting method proposed in Section~\ref{4.3}, we compare LALA with LALA-NB.
The loss function for LALA-NB's advisor is defined as (\ref{loss_Advisor}).
Compared with LALA's advisor loss (\ref{loss_Advisor_Boost}), LALA-NB does not involve the discriminator's judgement.
% Other details in LALA-NB are aligned with LALA-Boost.

% \item \textbf{LALA with state-action discriminator (LALA-SA)} - 
\subsubsection{LALA with state-action discriminator (LALA-SA)}
To verify the advantage of PLGAN proposed in Section~\ref{4.2}, we compare it with a state-action discriminator similar to MAGAIL \cite{song2018multi}.
The loss function for the discriminator is:
\begin{equation} \label{loss_Disc_SA}
\begin{aligned}
    \mathcal{L}_{Disc, i}^{\text{LALA-SA}} = - \mathbb{E}_{s} &\left[\log D^{\psi_i}(s, \pi^{\text{advisor}}) \right. \\
    & \left. + \log(1-D^{\psi_i}(s, \pi^{\text{agent}})) \right],
\end{aligned}
\end{equation}
where $\pi^{\text{agent}}$ is the same as (\ref{pi_agent}).
The loss function for each agent is defined as: 
\begin{equation}
    \mathcal{L}_{Agent, i}^{\text{LALA-SA}} = \mathcal{L}_i^{\theta_i}+\lambda  \mathbb{E}_{s} \log(1-D^{\psi_i}(s, \pi^{\text{agent}})).   
\end{equation}
The advisor's loss function is defined as:
\begin{equation}
\begin{aligned}
&    \mathcal{L}_{Advisor}^{\text{LALA-SA}} = \sum_{v \in \mathcal{V}} \left[ \sum _{u \in \mathcal{N}_t(v)} - \log (\sigma(\pi_v^T \pi_u)) - \right.\\
& \left. \sum _{w \in \mathcal{N}_s(v)} \log(\sigma(-\pi_v^T \pi_w)) + I_{label}(v)\cdot JSD(\pi_v || \pi_v^{\text{agent}})\right] \\
&- \mu \sum_{i=1}^N \mathbb{E}_{s} \left[\log D^{\psi_i}(s, \pi_v)\right].
\end{aligned}    
\end{equation}
Compared with the corresponding loss functions for LALA, i.e. (\ref{loss_Disc}), (\ref{loss_Agent}), and (\ref{loss_Advisor_Boost}), the input of the discriminator is changed from a set of state-action pairs to a single pair of state-action. 
Note that, unlike MAGAIL, LALA does not require demonstration data. The state data used to represent the advisor and agent is the same, while the action data is different.
Therefore, a set of sate-action pairs is more appropriate to manifest a policy than a single state-action pair, which suggests that LALA is expected to deliver more scalable and efficient performance than that of LALA-SA. 

% \end{itemize}

%%%%%%%%%%%%%%%%%%%%%%%%%%%%%%%%%%%%%%%%%%%%%%%%%%%%%%%%%%%%%%%%
\subsection{Configurations of LALA}
For the MARL agent module, each agent's decision is defined as (\ref{pi_agent}).
The network structure and training configurations are aligned with \cite{jin2021information}, which is the same as No-Advice.
Specifically, the agent module contains a Q-network that has two hidden layers containing $300$ and $200$ units, respectively. 
Following \cite{jin2021information}, an encoder is used to represent the prediction of other agents' behaviors, which has two hidden layers containing $32$ and $16$ units, respectively.
The replay buffer size is 1500, the batch size of SGD is 32.
Adam optimizer with a learning rate of 0.01 is applied to update the parameters.

In the DualGCN advisor module, we adopt the same structure for the temporal GCN and spatial GCN.
Specifically, the temporal/spatial aggregation function in (\ref{gcn}) is instantiated with a mean operator. 
Hidden layers $f^k$ and $g^k$ are fully connected layers with $N$ output units. The activation function $\phi$ adopts ReLU. The number of hidden layers is set as $K=2$.
Adam optimizer with a learning rate of 0.1 is applied to update the parameters.

For the discriminator, we adopt a three-layer Transformer encoder with four heads, 256 units, and 0.1 dropout rate in each layer.
Besides, the input layer projects the raw input as a 64-dimensional vector, and the output layer (classifier) projects the embedding features as a scalar activated by a sigmoid function.
Adam optimizer with a learning rate of 0.01 is applied to update the parameters.

%%%%%%%%%%%%%%%%%%%%%%%%%%%%%%%%%%%%%%%%%%%%%%%%%%%%%%%%%%%%%%%%%

\subsection{Macro-Level Performance}
    \subsubsection{Training performance}
\begin{figure}[t]
\centering
\subfigure
{
    \begin{minipage}[t]{4cm}
	\centering          
    \includegraphics[width=1.0\columnwidth]{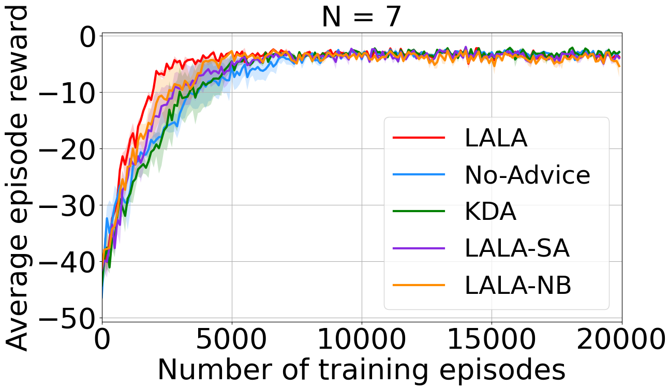}
    \end{minipage}
}
\subfigure
{
    \begin{minipage}[t]{4cm}
	\centering          
    \includegraphics[width=1.0\columnwidth]{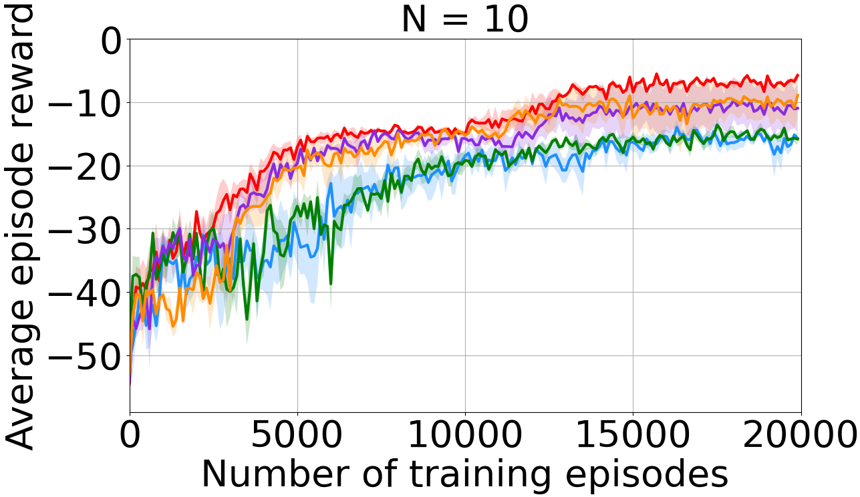}
    \end{minipage}
}
\caption {
Convergence curves of average episode reward.
}
\label{CovergenceCurves_reward} 
\end{figure}

\begin{figure}[t]
\centering
\subfigure
{
    \begin{minipage}[t]{4cm}
	\centering          
    \includegraphics[width=0.9\columnwidth]{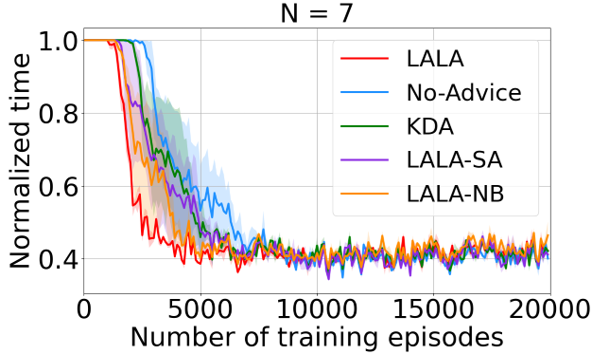}
    \end{minipage}
}
\subfigure
{
    \begin{minipage}[t]{4cm}
	\centering          
    \includegraphics[width=1.0\columnwidth]{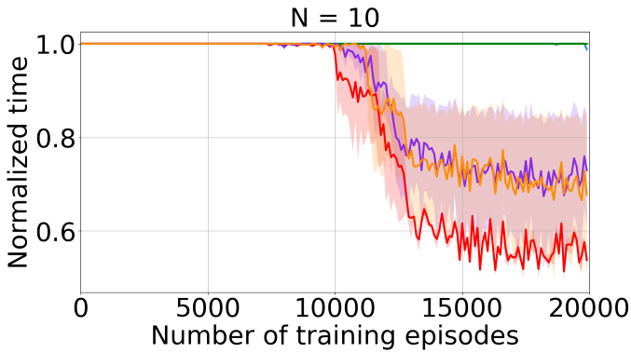}
    \end{minipage}
}
\caption {
Convergence curves of normalized navigation time.
}
\label{CovergenceCurves_time} 
\end{figure}

Fig.~\ref{CovergenceCurves_reward} shows the learning curves of the average episode reward of different methods in the cases of N=7 and N=10.
As can be seen from the results, LALA outperforms the other four methods consistently in terms of convergence speed.
Additionally, as N increases, LALA can get more average reward than other methods, and especially when N=10, the advantage is significant.
No-Advice performs the worst in terms of both convergence speed and average reward, as compared with the four methods involving advice.
It demonstrates that learning with advice can improve MARL.
Fig.~\ref{CovergenceCurves_time} shows learning curves of normalized navigation time of different methods.
The navigation time refers to the time used by reaching all targets, which equals the time taken by the agent who is the last one to reach a target. 
If an episode fails, the navigation time is recorded as the maximum episode length.
Results in Fig.~\ref{CovergenceCurves_time} indicate that LALA converges the most rapidly among all methods.

    \subsubsection{Execution performance}
We evaluate the execution performance of different methods with two metrics, i.e. cooperation success rate and normalized navigation time.
Successful cooperation means that the targets are reached respectively by different agents.
Namely, agents can avoid conflicts and achieve cooperation.
Table~\ref{tab:success_rate} and Table~\ref{tab:normalized_time} respectively show the test results of cooperation success rate and normalized navigation time.
Each result is recorded with the mean results of three random seeds used in training. 
As can be seen from the results, LALA achieves the highest success rate and the least navigation time among all methods.
Over different N values, LALA performs the most stably in all methods. 
Notably, the result corresponding to N=10 indicates the significant superiority of LALA over other methods. 

\begin{table}[thbp]
    % \footnotesize
    \centering
    \caption{Cooperation success rate of different methods}
    \label{tab:success_rate}
    \begin{tabular}{lccccc}
    \toprule [1pt]
               & LALA             & LALA-NB     & LALA-SA    & KDA     & No-Advice \\
    \hline
    N=7           & \textbf{0.930}   & 0.890       & 0.888      & 0.913   & 0.874    \\
    % N=8           & 0.847            & \textbf{0.862} & 0.775   & 0.583   & 0.547    \\
    % N=9           & 0.823            & \textbf{0.826} & 0.775   & 0.292   & 0.241    \\
    N=10          & \textbf{0.757}   & 0.520       & 0.449      & 0.000   & 0.015    \\
    % Average       & \textbf{0.839}   & 0.775       & 0.722      & 0.447   & 0.419    \\
    \bottomrule [1pt]
    \end{tabular}
\end{table}
\begin{table}[thbp]
\centering
    % \footnotesize
    \caption{Normalized navigation time of different methods}
    \label{tab:normalized_time}
    \begin{tabular}{lccccc}
    \toprule [1pt]
               & LALA             & LALA-NB     & LALA-SA    & KDA     & No-Advice \\
    \hline
    N=7           & \textbf{0.382}   & 0.408       & 0.402      & 0.392   & 0.414    \\
    % N=8           & 0.453            & \textbf{0.441} & 0.491   & 0.619   & 0.641    \\
    % N=9           & 0.470            & \textbf{0.469} & 0.507   & 0.812   & 0.843    \\
    N=10          & \textbf{0.516}   & 0.667       & 0.712      & 1.000   & 0.990    \\
    % Average       & \textbf{0.455}   & 0.496       & 0.528      & 0.706   & 0.722    \\
    \bottomrule [1pt]
    \end{tabular}
\end{table}
%%%%%%%%%%%%%%%%%%%%%%%%%%%%%%%%%%%%%%%%%%%%%%%%%%%%%%%%%%

\subsection{Micro-Level Information Gain}
\begin{figure}[t]
\centering
\subfigure
{
    \begin{minipage}[t]{4cm}
	\centering          
    \includegraphics[width=1.0\columnwidth]{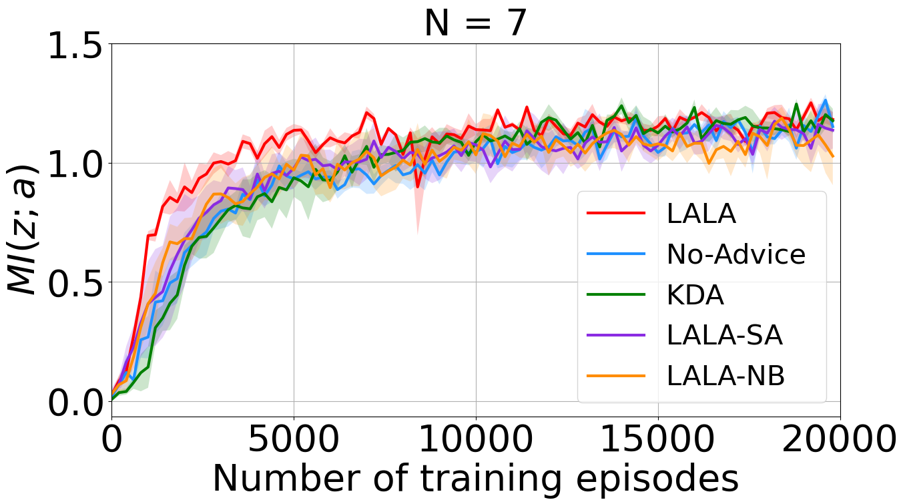}
    \end{minipage}
}
\subfigure
{
    \begin{minipage}[t]{4cm}
	\centering          
    \includegraphics[width=1.0\columnwidth]{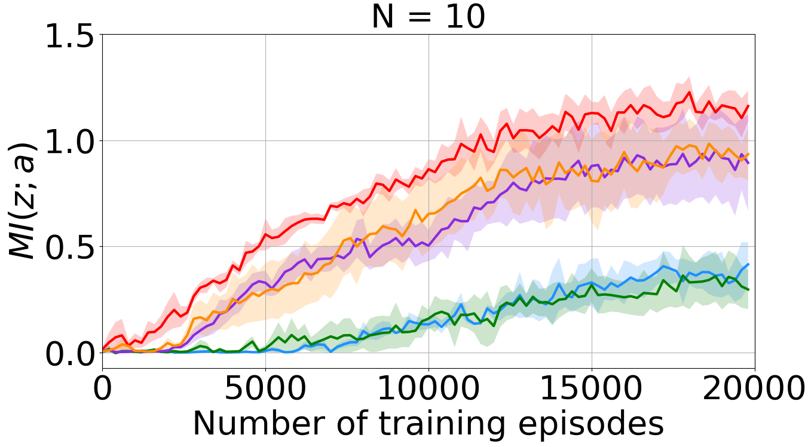}
    \end{minipage}
}
\caption {
Curves of mutual information gain.
}
\label{MI_curves} 
\end{figure}

To investigate what happens in agents' decision module at micro level during the process of learning to cooperate, we examine the mutual information between agent's local perceptions and other agents' actions.
Specifically, we look into the decision network of an agent, i.e. Fig.~\ref{iborm}, and compute the mutual information between the learned representation of other agents' actions and their true actions using the mutual information neural estimator (MINE) \cite{belghazi2018mutual}.
We denote the mutual information as $MI(z;a)$. 
The curves of the mutual information gain of different methods during training are shown in Fig.~\ref{MI_curves}.
As can be seen from the results, $MI(z;a)$ increases as training proceeds.
Remarkably, the variation of $MI(z;a)$ is positively correlated with the changes of average episode reward (Fig.~\ref{CovergenceCurves_reward}) in general.
Comparing LALA with No-advice, we can see that LALA can enhance $MI(z;a)$ by a large margin.
This phenomenon indicates that introducing meso-level coordination to advise micro-level policy learning can enhance agents' mutual understanding about their decisions.
Further, the enhancement of agents' mutual understanding promotes the macro-level cooperation performance.

%%%%%%%%%%%%%%%%%%%%%%%%%%%%%%%%%%%%%%%%%%%%%%%%%%%%%%%%%%%%%%%%%%%%%%%%

\subsection{Meso-Level Coordination}
\begin{figure}[t]
\centering
\subfigure
{
    \begin{minipage}[t]{3.7cm}
	\centering          
    \includegraphics[width=0.78\columnwidth]{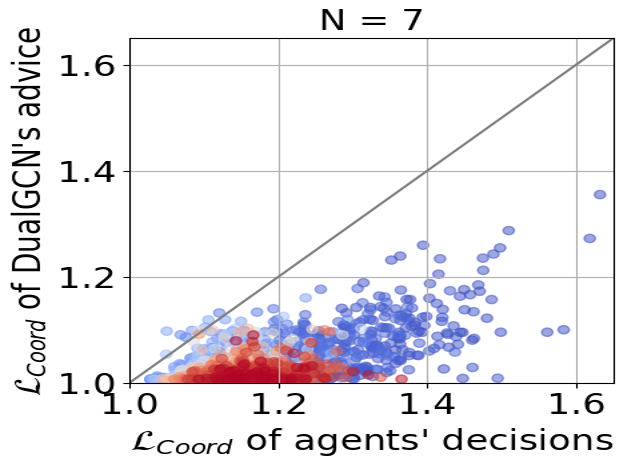}
    \end{minipage}
}
\subfigure
{
    \begin{minipage}[t]{4.3cm}
	\centering          
    \includegraphics[width=0.9\columnwidth]{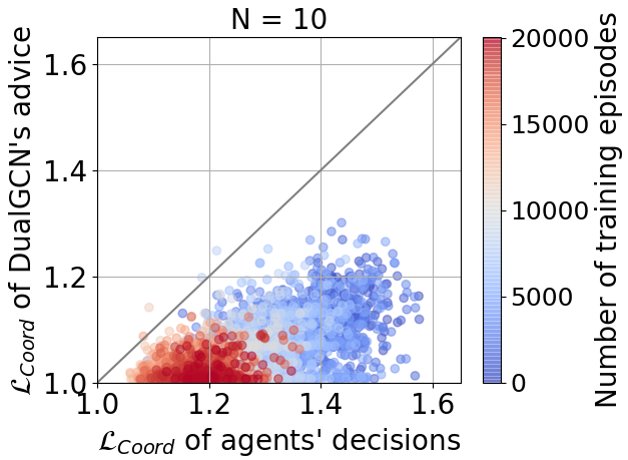}
    \end{minipage}
}
\caption {
Coordination loss of agents' decisions and DualGCN's advice.
}
\label{coloss} 
\end{figure}

To investigate the capability of DualGCN regarding meso-level coordination, we examine its cost terms that reflect the degree of conflict and temporal discontinuity of agents' decisions. 
We leverage the cost terms to examine the coordination of both the agents' decisions and the advice given by DualGCN.
To be specific, we define a coordination loss based on DualGCN's objective function (\ref{loss_Advisor_Boost}), which is given by:
\begin{equation}
\label{CoLoss}
% \footnotesize
\begin{aligned}
&    \mathcal{L}_{Coord} = \frac{1}{|\mathcal{V}|}\sum_{v \in \mathcal{V}} ( \frac{1}{|\mathcal{N}_t(v)|}\sum _{u \in \mathcal{N}_t(v)} - \log (\sigma(\pi_v^T \pi_u))\\
&\qquad \qquad \quad - \frac{1}{|\mathcal{N}_s(v)|} \sum _{w \in \mathcal{N}_s(v)} \log(\sigma(-\pi_v^T \pi_w)) ).
\end{aligned}
\end{equation}
We compute the coordination loss of DualGCN's input (agents' decisions) and output (advice) data in training. 
Fig.~\ref{coloss} shows the results of cases where N=7 and 10.
As can been seen from the results, the coordination loss of DualGCN's advice is lower than that of agents' decisions, which demonstrates the coordination capability of DualGCN.
As the number of training episodes increases (from blue to red), the coordination loss of both agents' decisions and DualGCN's advice decreases, which indicates the coordination improvement in both micro-level agents and meso-level DualGCN as training proceeds.

\subsubsection{Effect of advisor booster}
\begin{figure}[t]
\centering
\subfigure
{
    \begin{minipage}[t]{4cm}
	\centering          
    \includegraphics[width=1\columnwidth]{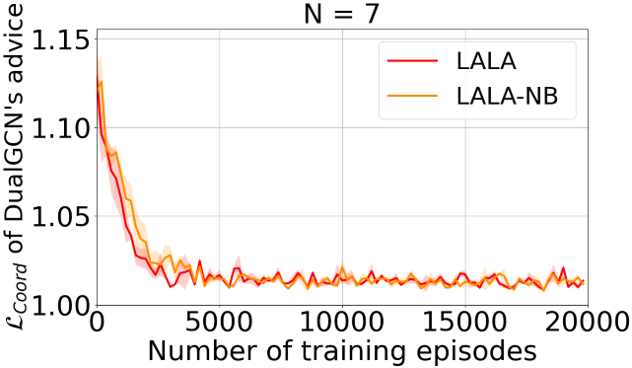}
    \end{minipage}
}
\subfigure
{
    \begin{minipage}[t]{4cm}
	\centering          
    \includegraphics[width=1\columnwidth]{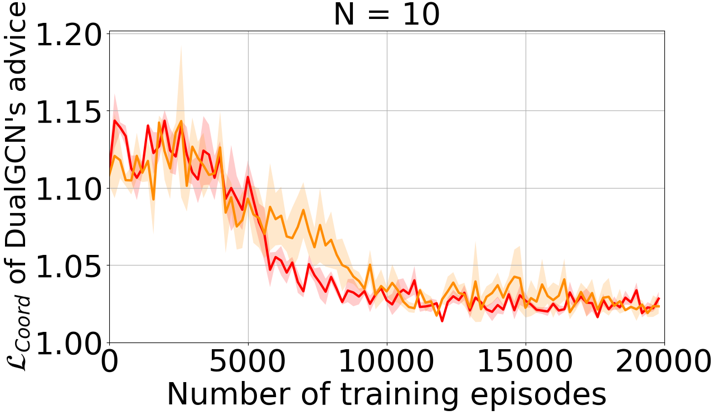}
    \end{minipage}
}
\caption {
Convergence curves of DualGCN's coordination loss. 
% of LALA and LALA-NB.
}
\label{coloss_lalaVSnb} 
\end{figure}

To investigate how the advisor booster affects the coordination capability of DualGCN, we compare the convergence curves of the coordination loss of DualGCN trained with LALA and LALA-NB.
Results are presented in Fig.~\ref{coloss_lalaVSnb}.
Compared with LALA-NB, LALA can promote DualGCN to achieve lower coordination loss during training. 
It verifies that the advisor booster based on the discriminator's discerning power can further enhance the coordination capability of DualGCN. 
Besides, for each curve, the convergence point appears earlier than the macro-level average episode reward (Fig.~\ref{CovergenceCurves_reward}).
It indicates that meso-level coordination advising causes the improvement of macro-level cooperation performance. Due to the time cost by learning from the advice at micro level, the emergence of macro-level cooperation is delayed. 

\subsubsection{Effect of DualGCN's depth}
\begin{figure}[t]
\centering
\subfigure
{
    \begin{minipage}[t]{4cm}
	\centering          
    \includegraphics[width=1\columnwidth]{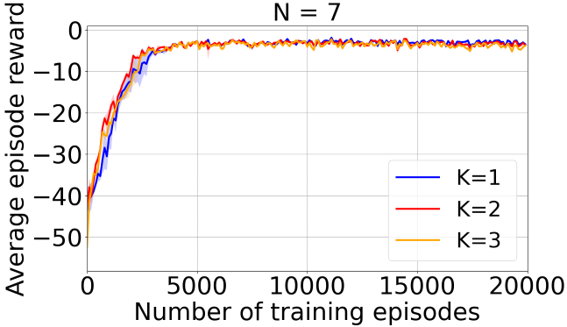}
    \end{minipage}
}
\subfigure
{
    \begin{minipage}[t]{4cm}
	\centering          
    \includegraphics[width=1\columnwidth]{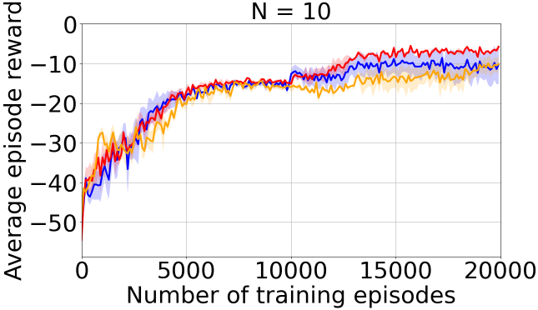}
    \end{minipage}
}

\subfigure
{
    \begin{minipage}[t]{4cm}
	\centering          
    \includegraphics[width=1\columnwidth]{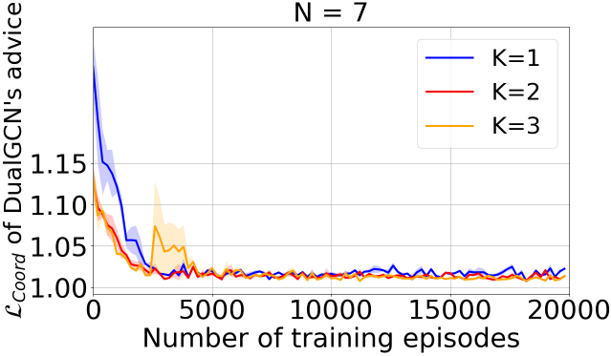}
    \end{minipage}
}
\subfigure
{
    \begin{minipage}[t]{4cm}
	\centering          
    \includegraphics[width=1\columnwidth]{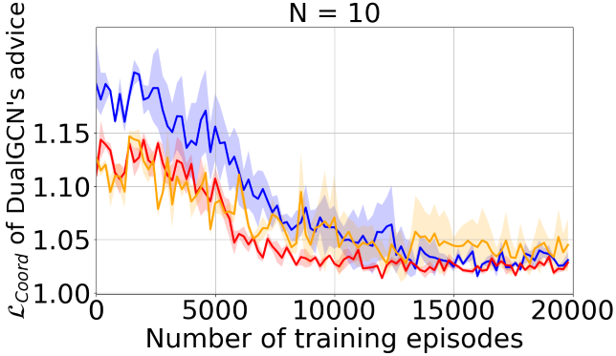}
    \end{minipage}
}
\caption {
Performance comparison of different depths of DualGCN. Top row: average episode reward. Bottom row: coordination loss.
% of LALA and LALA-NB.
}
\label{gcnlayer} 
\end{figure}

The depth of DualGCN determines the scope of meso-level coordination of LALA.
In the above experiments, the number of hidden layers is K=2. 
In this part, we compare the performance of LALA with K=1, 2, and 3.
Results are shown in Fig.~\ref{gcnlayer}.
As can be seen from the bottom row of Fig.~\ref{gcnlayer}, K=1 results in more coordination loss, which implies the degraded coordination capability.
The degraded coordination capability causes less reward, as shown in the top row of Fig.~\ref{gcnlayer}. 
On the other side, when K=3, the coordination capability also degenerates. A possible reason is that when K increases, the learned features of each vertex tend to be similar and thus hinder the resolution of conflicts.

%%%%%%%%%%%%%%%%%%%%%%%%%%%%%%%%%%%%%%%%%%%%%%%%%%%%%%%%%%%%%%%%%%%%%%%%
\section{Conclusion and future work}
We propose LALA approach from the perspective of multilevel emergence dynamics, which leverages the meso-level decision interaction based on agents' spatiotemporal decision structure to advise micro-level decision-making, and thus propels the emergence of macro-level cooperation. 
LALA involves three-part learning, i.e., DualGCN advisor, agents' decision networks, and advisor-agent discriminators.
The DualGCN aggregates agents' decision information and mitigates decision conflicts and discontinuity in spatial and temporal domains, respectively.
Agents learn from the coordinated decision advice generated by the DualGCN via PLGAN in addition to their own explorations.
The PLGAN includes a discriminator which distinguishes policies of the agent and the advisor, and boosts both of them via its judgement.

% Experimental results demonstrate the coordination capability of our proposed DualGCN advisor.
By examining the mutual information between agent's local perception and other agents' actions, we reveal that the meso-level coordinated decision advice helps agents' mutual understanding regarding their decisions.
Experimental results demonstrate that LALA's advantage in macro-level cooperation performance is positively correlated with its advantage in micro-level mutual information gain and meso-level coordination capability.

Our work elucidated the importance of meso-level coordination for cooperative MARL. 
We verified our method in a cooperative navigation task.
Although it is a representative and complex multi-agent task, the structure of meso-level coordination is limited  relatively.
As a future work, we will verify LALA in other tasks to reveal more inspiring results.
In addition, the experimental results imply the potential of developing a more powerful GCN and discriminator to promote meso-level coordination, which is worthy of further studies.

\bibliographystyle{IEEEtran}
\bibliography{IEEEabrv,main}

\end{document}